# Direct measurement of the viscocapillary lift force near a liquid interface

Hao Zhang[1], Zaicheng Zhang[1,2], Aditya Jha[1], Yacine Amarouchene[1], Thomas Salez[1], Thomas Guérin[1], Chaouqi Misbah[3] and Abdelhamid Maali[1,*]

[1] Université de Bordeaux & CNRS, LOMA, UMR 5798, F-33400 Talence, France
[2] School of Physics, Beihang University, 100191, Beijing, China
[3] Université Grenoble Alpes, CNRS, LIPHY, F-38000 Grenoble, France

Abstract:

Lift force of viscous origin is widespread across disciplines, from mechanics to biology. Here, we present the first direct measurement of the lift force acting on a particle moving in a viscous fluid along the liquid interface that separates two liquids. The force arises from the coupling between the viscous flow induced by the particle motion and the capillary deformation of the interface. The measurements show that the lift force increases as the distance between the sphere and the interface decreases, reaching saturation at small distances. The experimental results are in good agreement with the model and numerical calculation developed within the framework of the soft lubrication theory.

* Corresponding author: abdelhamid.maali@u-bordeaux.fr

A lubricated motion between two sliding objects is encountered in numerous scenarios spanning various length scales. Examples include cellular motion in blood vessels [1, 2], skiers on snow [3], and tires that experience aquaplaning [4].

At low Reynolds numbers, the reversibility of hydrodynamic equations typically precludes the existence of a normal force on a sphere sliding in fluid along a rigid wall. The pressure field in this scenario displays antisymmetry: it is positive in the direction of motion and negative in the opposite direction, resulting in a null net normal force. However, when the wall is soft, the viscous pressure within the confined lubricant induces deformation of the wall, disrupting the antisymmetry of the pressure field and engendering a repulsive force [5-10].

Elastohydrodynamic (EHD) coupling refers to the coupling between the hydrodynamic viscous pressure and the elastic deformation of the sample [5-26]. Central to this coupling is the softness parameter, contingent upon the ratio between the hydrodynamic pressure and the elastic modulus of the materials involved. In scenarios where surfaces are rigid and the softness parameter approaches zero (thus rendering the EHD effect negligible), no normal force is discernible. Conversely, for soft surfaces characterized by a substantial softness parameter, a notable normal force can be induced, and its intensity increases as the softness parameter increases.

Theoretical studies have dealt with the lift force exerted on a sphere sliding over a soft solid, exploring scenarios with diverse sample thicknesses and material properties, spanning from compressible to incompressible materials and viscoelastic fluids [5-10, 23, 24]. Further studies have analyzed the involvement of the lift force in the motion of biological cells and vesicles (representing a biomimetic cell membrane) subjected to shear flow within confined channels [26-29]. The calculations suggest a compelling correlation: the softer the membrane of the cells, the greater the lift force, and correspondingly, the higher the shear rate, the more pronounced the lift force becomes.

Experiments on lift force have been conducted recently using various techniques, focusing on spheres and cylinders sliding on soft solid surfaces as well as on elastic membrane [25, 30-33].

The interface between immiscible liquids offers another intriguing system where a repulsive lift force can emerge when a sphere moves in proximity to the interface. This phenomenon involves the intricate interplay between the viscous pressure and the interface itself, as described by the Young-Laplace equation. The resultant lift force, stemming from hydro-capillary coupling, holds significance in diverse realms. For instance, it could influence the dynamics of droplets and bubbles within microfluidic systems [34] and potentially shape the behavior of active particles and self-propelled objects near interfaces [35]. However, it is noteworthy that, to the best of our knowledge, this force has not been measured directly and quantitatively.

In this letter, we present a direct measurement of the lift force acting on a sphere moving in a viscous fluid along the liquid interface that separates two liquids. Using an atomic force microscope (AFM), the viscocapillary lift force was measured versus the distance between the sphere and the interface. We investigated various liquid interfaces, working frequencies, sliding velocities, and two different sphere radii. Our measurement results are in good agreement with the theoretical model and numerical calculation developed within the framework of soft lubrication theory.

A schematic of the experimental setup is shown in **Fig. 1**. An AFM cantilever affixed with a rigid sphere was employed to measure the viscocapillary lift force near the liquid-liquid interface, and the relative lateral velocity of the interface with respect to the sphere is induced by a piezo stage. The experiment was performed using an AFM (Bioscope, Bruker) equipped with a cantilever holder (DTFML-DD-HE) that allows working in a liquid environment. Two spherical borosilicate particles (MO-Sci Corporation) with radii of $R = 53 \pm 1$ $\mu$m and $36 \pm 1$ $\mu$m were used. The spheres were thoroughly cleaned with ethanol and pure water before being glued to the end of a silicon nitride triangular-shaped cantilever (Model: SNL-10, Bruker) using epoxy (Araldite, Bostik, Coubert). The interface was formed by depositing a drop of glycerol ($\geq$ 99.5%, Sigma-Aldrich) onto a mica surface, followed by the deposition of a second drop of silicone oil on top of the glycerol drop. Two silicone oils (Silicone oil AR 20 and Silicone oil AS 100, Aldrich) with viscosity of 20 mPa·s and 100 mPa·s were investigated. The mica substrate, on which the interface was fixed, was mounted on a multi-axis piezo system (NanoT series, Mad City Labs). The piezo system enables the induction of a shear flow by displacing the sample laterally in the direction perpendicular to the cantilever beam. Additionally, it facilitates the monitoring of the distance $d$ between the sphere and the sample by vertically displacing the sample.

The stiffness of the cantilever was obtained using the drainage method as described in ref [36]. From the drainage experiment conducted on a mica substrate (without liquids interface), the measured values of the cantilever stiffness with an attached sphere were $k_c = 0.16 \pm 0.02$ N/m and $k_c = 0.20 \pm 0.02$ N/m for spheres with radii of $R = 36$ $\mu$m and $R = 53$ $\mu$m, respectively. The interfacial tension was measured using the pendant drop method and analyzed using the *ImageJ* plugin [37], yielding a value of $\sigma = 21 \pm 2$ mN/m for both the silicone oil 20-glycerol interface and the silicone oil 100-glycerol interface.

A method that allows for obtaining large lateral velocities and compatible with colloidal AFM probe setups is to oscillate the substrate laterally. The lateral velocity amplitude $A\omega$ can be monitored by varying either the radial frequency or the amplitude of the substrate oscillation in the lateral direction. In our experiments, we employed two working frequencies, 5 Hz and 10 Hz, while controlling the amplitudes $A$ to result in lateral amplitude velocities ranging between 1.22 and 3.06 mm/s. The expression of the lift force for a sphere in relative motion near (and parallel to) a fluid surface can be obtained for small Reynolds number in the regime of lubricated flow ($d \ll R$). We denote $x$ and $y$ as the spatial coordinates in the plane of the interface, with the origin at the center of the sphere. We consider that the fluid velocity vanishes at the surface of the sphere, while at the interface with the lower fluid, and we assume that the viscosity of the lower fluid is sufficiently high such that the velocity at the lower surface is $v\hat{e}_x$. With these assumptions, in the framework of the lubrication approximation, the pressure $p$ in the gap is described by the Reynolds equation [10]:

$$\nabla \left(\frac{h^3}{6} \nabla p\right) + v \nabla h \cdot \hat{e}_x = 0 \quad (1)$$

$\nabla = \partial_x \hat{e}_x + \partial_y \hat{e}_x$ is the two-dimensional nabla operator, and $h$ is the gap thickness given by:

$$h = h_0(r) + \delta \quad (2.a)$$

$$h_0(r) = d + \frac{x^2 + y^2}{2R} \quad (2.b)$$

where $\delta$ is the deformation of the interface and $r = \sqrt{x^2 + y^2}$ is the distance to center of the sphere. In the lubrication approximation, if the viscosity of the lower fluid is sufficiently high to consider that the pressure is not modified inside it, the Young-Laplace equation relates $\delta$ to the pressure:

$$p + \sigma \nabla^2 \delta = 0 \quad (3)$$

where $\sigma$ is the surface tension of the interface between the two liquids and $\nabla^2$ is the two-dimensional Laplacian operator.

For a non-deformable interface $\delta = 0$ and the pressure field is given by [38]:

$$p_0(x, y) = \frac{6\eta v x}{5 h_0^2} \quad (4)$$

where $\eta$ is the viscosity of the upper fluid.

Interface deformation leads to the modification of the local gap distance, consequently affecting the pressure field. Assuming small variation in pressure, at the leading order, we derive the force acting on the sphere in the vertical direction as (see **Supplemental Material** [39]):

$$F_{lift} = \frac{6\pi}{25} \frac{\eta^2 v^2 R^3}{\sigma d^2} \quad (5)$$

**Equation (5)** shows that the lift force has a quadratic dependence on the velocity and viscosity. The force increases with the cubic size of the sphere and is inversely proportional to the value of the surface tension of the interface, and is inversely proportional to the square of the distance. In the experiments, the velocity oscillates as $v(t) = A\omega cos(\omega t)$, so that the force can be expressed as the sum of a time independent term $3\pi\eta^2 A^2 \omega^2 R^3/(25\sigma d^2)$ and a time dependent term $3\pi\eta^2 A^2 \omega^2 R^3 cos(2\omega t)/(25\sigma d^2)$ oscillating with a radial frequency of $2\omega$. In our measurements, we focus only on the time-independent term, obtained by measuring the time-averaged force applied on the sphere. Note that, the prefactor for the time-averaged force is half of that in **Eq. (5)**.

**Figure 2** presents the measured time-average force $F = <F_{lift}>$ for a sphere immersed in silicone oil 20, sliding along the interface between silicone oil and glycerol, and versus the distance between the sphere and the interface. It also presents the force measured with the same sphere, the same working frequency, and the same amplitude immersed in silicone oil 20 but sliding along a rigid mica substrate. It is notable that, as expected, there is no repelling lift force observed on the mica substrate, where the deformation is negligible due to the high rigidity of the mica. However, near the soft liquid-liquid interface, we observe a force that increases as the gap distance is reduced.

In **Fig. 3(a)**, force measurements at varying velocities on the silicone oil 20-glycerol interface are presented, showing an increasing trend with the amplitude velocity. The force divided by the square of the velocity is shown in the inset of **Fig. 3(a)**. As expected from the expression of the lift force, the values of the normalized forces coincide with each other. **Figure. 3(b)** shows the force measurements using silicone oils of viscosities 20 and 100 mPa·s. The force increases noticeably as the viscosity increases. It is worth noting that the interfacial tension remains consistent between silicone oil 20-glycerol and silicone oil 100-glycerol interfaces, with only viscosity differing. The normalized force, obtained by dividing by the square of viscosity, coincides between measurements conducted with both types of silicone oil, which is consistent with the prediction of **Eq. (5)**. **Figure. 3(c)** shows the force measurements using different sphere sizes are illustrated. Larger sphere results in higher force, as expected. The inset graph shows the normalized forces relative to the cube of the radius, demonstrating agreement across sphere sizes. Furthermore, as depicted in the insets of **Fig. 3**, the log-log plots indicate that the lift force is proportional to $d^{-2}$ at large distances. The measurements suggest a clear correspondence between the lift force and the relevant parameters, consistent with the expression in **Eq. (5)**.

To rationalize our measurements and to comprehensively understand the dependence of the lift force on parameters such as distance, sphere size, viscosity, and velocity, we introduce the dimensionless softness parameter [6] $\kappa = \eta A \omega R^{\frac{3}{2}}/\left(\sigma d^{\frac{3}{2}}\right)$ and the **Eq. (5)** can be rewritten as the normalized force $F/F^*$:

$$\frac{F}{F^*} = \left(\frac{3\pi}{25}\right)\kappa \qquad (6)$$

where $F^* = \eta A \omega R^{\frac{3}{2}}/d^{\frac{1}{2}}$.

The measurements obtained with different working frequencies, amplitude velocity and viscosity as well as the sphere sizes are shown in **Fig. 4**. At very small softness parameters corresponding to the large distance regime, all the measurements coincide with each other, and they scale linearly with the softness parameter, as predicted by **Eq. (6)** (shown as the continuous red curve in **Fig. 4**). At higher values of $\kappa$, the experimental curves $F/F^*$ show a saturation with the softness parameter, with small deviations that increase with increasing radial frequency, viscosity and sphere size. In the limit of low excitation frequency, the saturation can be predicted by considering that the deformation of the interface is no longer linear with the velocity. Indeed, the numerical integration of Reynolds (**Eq. (1)**) and Young-Laplace equations (**Eq. (3)**) leads to the black curve in **Fig. 4**, which shows a saturation at high softness values. The results show a good agreement between the calculations and experiments, but small deviations were observed at large values of $\kappa$. The rotation of the sphere induced by the lubricated flow could be invoked to explain the deviation between the calculation and the measurements. In our setup, the sphere is fixed on the cantilever, and as the gap is reduced, the hydrodynamic torque induces a rotation of the sphere, which reduces the relative velocity between the interface and the sphere. From the measured lateral torsion signal of the cantilever, we extracted the rotational velocity of the sphere and found it to be very small (less than 1% of the value of the sliding velocity). Therefore, it cannot explain the deviation at large values of $\kappa$ between the measurements and the calculations.

We observe that at a given viscosity, the experimental curve at the higher excitation frequency deviates significantly from the theoretical curve, which indicates that non-stationary effects could be responsible for the non-collapsing of the experimental curves at large values of $\kappa$. To investigate the non-stationary effects, we relax the hypothesis of constant velocity $v$, so that the interface profile becomes time dependent, and the boundary condition is that the vertical component of the fluid flow at the interface is $-\partial_t \delta$. The equations become:

$$\frac{s}{\kappa^{1/3}} \partial_\tau \Delta = \nabla_\parallel \left(\frac{H^3}{12} \nabla_\parallel P\right) + \frac{\cos(\tau)}{2} \nabla_\parallel H \cdot \hat{\boldsymbol{e}}_x \qquad (7)$$

where $s = (R^3 \omega \eta/(A^2 \sigma))^{\frac{1}{3}}$, and we have used the dimensionless times $\tau = \omega t$, pressure $P = p\, d^{\frac{2}{3}}/(\eta A \omega \sqrt{R})$, height $H = h/d$, lengths $X = x/\sqrt{Rd}$, $Y = y/\sqrt{Rd}$, and $\nabla_\parallel = \partial_X \hat{e}_x + \partial_Y \hat{e}_x$. Each curve in **Fig. 4**, where $d$ is varied while other parameters are kept constant, corresponds to a constant value of $s$. Numerical integration of the time-dependent partial differential equations leads to curves showing at high $\kappa$ a saturation value that depends on the value of $s$, as observed qualitatively in the experiments. However, the agreement is not quantitative, the theoretical predictions using a value of $s = 0.1$ is very close from those obtained for the static case.

The disagreement between theory and experiment at large $\kappa$, could be explained by considering the finite viscosity of the lower fluid which was, up to now, considered to be infinite. The finite viscosity of the lower liquid (glycerol) leads to the decreases of the velocity of the interface from the imposed value $v$ by the displacement of the mica substrate. The velocity at the interface can be obtained by equating the lateral shear stress on both sides of the interface. On the upper side, the stress is $\sigma_{xx} \sim \eta v_2/d$, and on the lower side it is $\eta_0 (v - v_2)/l$, where $\eta_0$ is the viscosity of glycerol and $l$ is the characteristic length scale of the flow, approximately a few $\sqrt{2Rd}$. From these equal quantities, we obtain the value of the interface velocity: $v_2 \sim v/(1 + \eta l/\eta_0 d)$. Hence, the interface velocity is close to the imposed velocity ($v_2 \approx v$) only at large distances $d \gg \eta l/\eta_0$. This condition can also be formulated as a condition on the softness parameters (see **Supplemental Material** [39]): the lower fluid viscosity is expected to be important when $\kappa \gg \kappa^*$, with $\kappa^* \sim \eta_0^3 v/(\eta^2 \sigma)$. In our experiments, typical values of $\kappa^*$ are of the order of a few tens, but estimating the actual value of the softness parameter require a precise estimation of the length scale at which the flow develops in the lower fluid, which could be larger than $\sqrt{2Rd}$. Understanding non-stationary effects with a lower fluid of finite viscosity is beyond the scope of this work, but we believe that they could be responsible for the small discrepancy between theory and experiments at large softness parameter.

In conclusion, our experiments provide a direct quantitative measurement of the lift force acting on a sphere moving along liquid-liquid interface. Through systematic investigation of the key parameters that control the lift force, we revealed the important trends governing the behavior of the lift force. At very small softness parameters, the normalized force of all measurements coincides with each other, and displays a linear scaling with the softness parameter as predicted by our model. However, as the softness parameter increases, saturation occurs, which is consistent with numerical calculations, and small discrepancies emerge owing to non-stationary effects and finite viscosity of the lower liquid. This study contributes to a deeper understanding of interfacial phenomena and paves the way for further study of the motion of objects close to liquid-liquid interface. In the future, including the non-stationary effect due to the time dependent velocity may help to explore the small discrepancy with the numerical calculations observed at saturation.

# Acknowledgements

Hao Zhang acknowledges financial support from the China Scholarship Council. The authors acknowledge the French National Research Agency through the supporting grants EDDL (ANR-19-CE30-0012), AMARHEO (ANR-18-CE06-0009-01), EMetBrown (ANR-21-ERCC-0010-01), Softer (ANR-21-CE06-0029) and Fricolas (ANR-21-CE06-0039). The authors also acknowledge financial support from the European Union through the European Research Council under an EMetBrown (ERC-CoG-101039103) grant. Views and opinions expressed are, however, those of the authors only, and do not necessarily reflect those of the European Union or the European Research Council. Neither the European Union nor the granting authority can be held responsible for them.

**Figures**

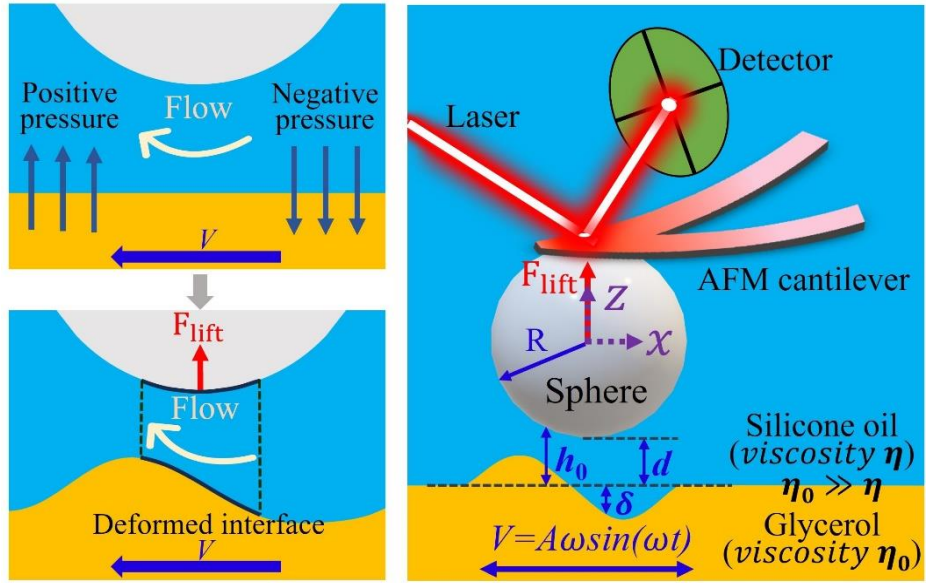

**FIG.1.** Sketch of the measurement of the lift force generated by the relative sliding of a rigid sphere and a liquid-liquid interface using an AFM-based experimental setup. A rigid borosilicate sphere is glued on to the end of an AFM cantilever beam. The soft interface is formed between the upper silicone oil liquid and the lower glycerol liquid. The liquids are deposited on solid substrate fixed on a piezo stage capable of lateral oscillation. The flow generated by the motion induces deformation of the interface. The lift force exerted on the sphere is directly measured from the vertical deflection of the AFM cantilever.

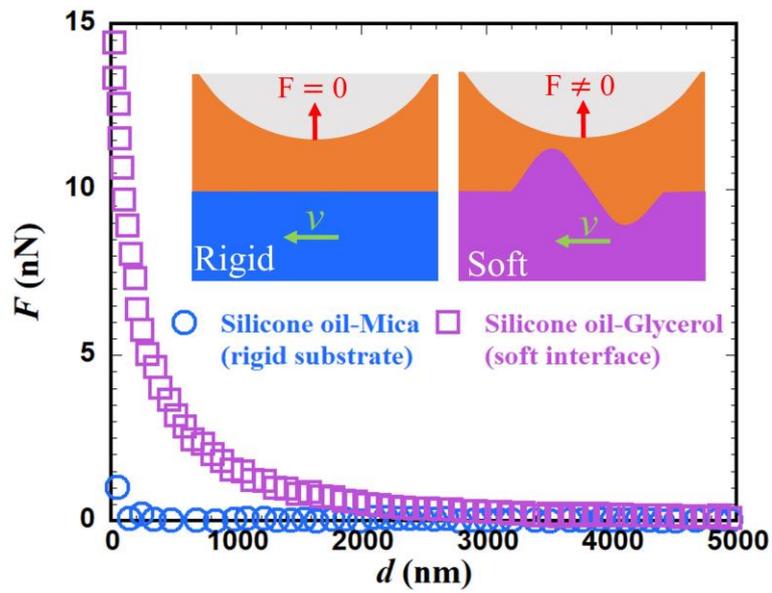

**FIG. 2.** The measured force on the soft silicone oil-glycerol interface (square) and on the rigid mica substrate (circle). Both measurements were conducted at a frequency of 10 Hz and a lateral velocity amplitude ($A\omega$) of 2.59 mm/s, using a sphere with a radius of 36 $\mu$m and silicone oil viscosity of 20 mPa·s.

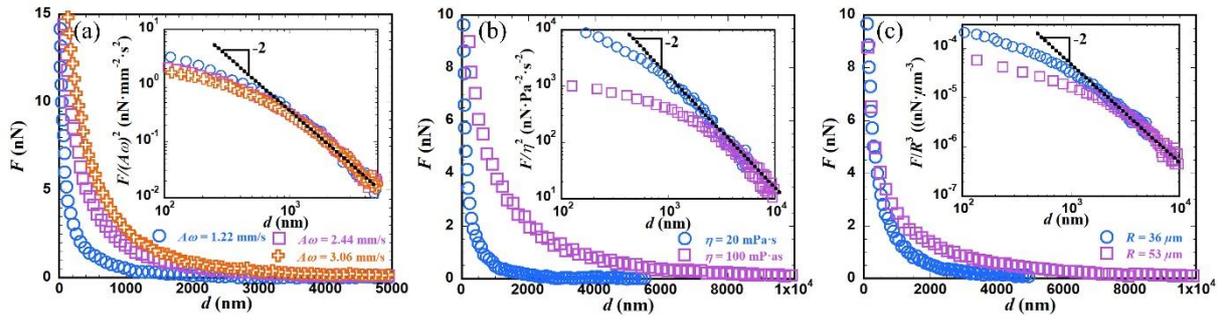

**FIG. 3.** The force measured is presented as a function of the gap distance to the silicone oil-glycerol interface for various sets of parameters. The insets display the normalized forces versus distance on logarithmic scales, with dashed lines in all the inset figures indicating a power law of -2. (a) Three different amplitude velocities with the same frequency are investigated. The inset illustrates the force divided by the square of the velocity. The viscosity of silicone oil is 20 mPa·s, the radius of the sphere is 36 $\mu$m, and the working frequency is 5 Hz. (b) Two different viscosities (20 and 100 mPa·s) are investigated. The inset shows the force divided by the square of the viscosity. The radius of the sphere is 36 µm, and the velocity amplitude is 1.22 mm/s with a working frequency of 5 Hz for both measurements. (c) Two different radii of the sphere are investigated. The inset shows the force divided by the cube of the radius. The viscosity of silicone oil is 20 mPa·s, the velocity amplitude is 2.59 mm/s with a working frequency of 10 Hz for both measurements.

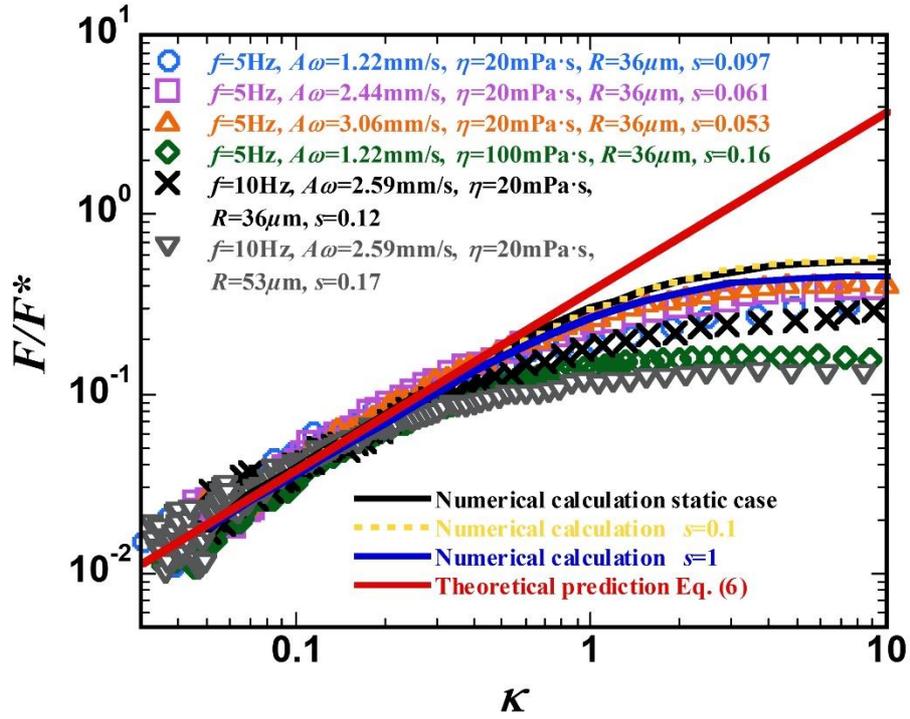

**FIG. 4**. The dimensionless force $F/F^*$ is plotted as a function of dimensionless compliance $\kappa$ on logarithmic scales. The hollow points, distinguished by different shapes and colors, represent measurements conducted with various parameters, including velocity amplitude, working frequency, radius of the sphere, and viscosity of the upper liquid. The continuous red line corresponds to the theoretical prediction obtained from **Eq. (6)**, while the black, yellow, and blue curves indicate numerical calculation results for $s$ values of 0.01 (static), 0.1, and 1, respectively.